\documentclass{article}

\usepackage{PRIMEarxiv}

\usepackage[utf8]{inputenc} 
\usepackage[T1]{fontenc}    
\usepackage{hyperref}       
\usepackage{url}            
\usepackage{booktabs}       
\usepackage{amsfonts}       
\usepackage{nicefrac}       
\usepackage{microtype}      
\usepackage{lipsum}
\usepackage{fancyhdr}       
\usepackage{graphicx}       
\graphicspath{{media/}}     

\usepackage{subcaption}
\usepackage{amsmath}
\usepackage[usenames,dvipsnames]{color}

\title{Presenting a Larger Up-to-date Movie Dataset and Investigating the Effects of Pre-released Attributes on Gross Revenue
\thanks{\textit{\underline{Citation}}: 
\textbf{
    Sharma, A. S., Roy, T., Rifat, S. A. \& Mridul, M. A. (2021). Presenting a Larger Up-to-Date Movie Dataset and Investigating the Effects of Pre-Released Attributes on Gross Revenue. 
    Journal of Computer Science, 17(10), 870-888. DOI:10.3844/jcssp.2021.870.888
}} 
}

\author{
    Arnab Sen Sharma, 
    Tirtha Roy,
    Sadique Ahmmod Rifat,
    Maruf Ahmed Mridul\\
  Department of Computer Science and Engineering\\
  Shahjalal University of Science and Technology \\
  Sylhet-3114, Bangladesh\\
  \texttt{
    arnab-cse@sust.edu, 
    tirtharoy1999@gmail.com,
    sadiqueahmmodrifat@gmail.com,
    mridul-cse@sust.edu} \\
}

\begin{document}
\maketitle

\begin{abstract}
Movie-making has become one of the most costly and risky endeavors in the entertainment industry. Continuous change in the preference of the audience makes it harder to predict what kind of movie will be financially successful at the box office. So, it is no wonder that cautious, intelligent stakeholders and large production houses will always want to know the probable revenue that will be generated by a movie before making an investment.  Researchers have been working on finding an optimal strategy to help investors in making the right decisions. But the lack of a large, up-to-date dataset makes their work harder. 
In this work, we introduce an up-to-date, richer, and larger dataset that we have prepared by scraping \href{https://www.imdb.com}{IMDb} for researchers and data analysts to work with. The compiled dataset contains the \textit{summery} data of 7.5 million titles and \textit{detail} information of more than 200K movies. Additionally, we perform different statistical analysis approaches on our dataset to find out how a movie's revenue is affected by different pre-released attributes such as budget, runtime, release month, content rating,  genre etc. In our analysis we have found that having a star cast/director has a positive impact on generated revenue.  
We introduce a novel approach for calculating the star power of a movie. Based on our analysis we select a set of attributes as features and train different machine learning algorithms to predict a movie's expected revenue. Based on generated revenue, we classified the movies in 10 categories and achieved a one-class-away accuracy rate of almost 60\% (bingo accuracy of 30\%). All the generated datasets and analysis codes are available online. We also made the source codes of our scraper bots public, so that researchers interested in extending this work can easily modify these bots as they need and prepare their own up-to-date datasets.
\end{abstract}


\section{Introduction}
     Among the multiple branches of the mammoth entertainment industry, motion pictures is one of the few largest and most booming multi-billion dollar industries; often reflecting major world events and promoting different cultures around the world. In the last few decades, the world has seen remarkable growth in the movie industry. Investors play a key role in this rapid growth. The primary purpose of making a movie from the investors’ and producers’ point of view is to to maximize the profit. The gross revenue of a movie mainly comes from theatrical earnings. But it can generate revenues from other platforms such as different online streaming platforms, DVDs, television broadcast rights, and merchandising. According to \textit{MPAA} (Motion Pictures Association of America), in the year 2012, the box office of the United States and Canada had a total revenue of over 10.8 Billion USD. \textit{Investopedia} stated that in order to make a major studio movie, the average production cost is roughly 65 million USD.

\begin{figure}
\centering
    \includegraphics[totalheight=6cm]{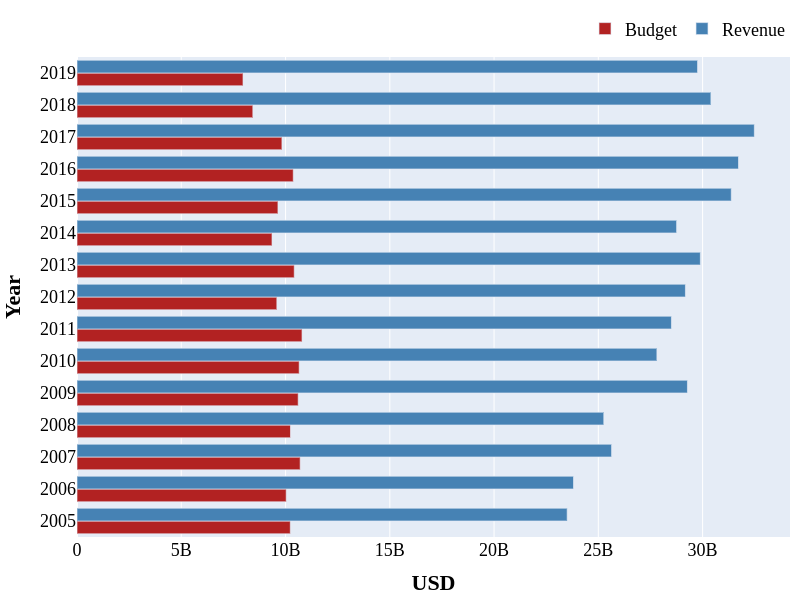}
    \caption{Year-wise total budget and revenue}
    \label{fig: bud_rev_year}
\end{figure}
    
    In \textbf{\color{BlueViolet}Figure \ref{fig: bud_rev_year}}, we have plotted the year-wise total revenue and investment/budget in movies in our \textit{cleaned} dataset from the year 2005 to 2019. From this figure, it can be observed that the total revenue always far exceeds the total budget for a year. But this figure can be misleading as the revenue distribution of box-office is not \textit{normal}. Instead of following a \textit{normal} distribution or \textit{Bell Curve}, it follows a \textit{Lévy} Distribution, where the distribution is heavily skewed to the right. The revenue values in the distribution are dominated by a few outlier blockbuster movies located in the far right tail. In fact, among movies released between the year 2000 and 2010 in the US, only 36\% produced a revenue higher than the production budget [\cite{lash2016early}]. 
    
    There is no doubt that a huge budget ensures better cast, better location, costly marketing strategies, use of modern technologies (like CGI) etc.; and these elements can play a vital role in a movie’s financial success. However, the  highly anticipated, huge budget (159 million USD) 2019 film ‘The Irishman’,  directed by the famous director Martin Scorsese, featuring the famous acting duo Robert De Niro and Al Pacino, fell surprisingly short of it’s expectations, even though the actors worked with the said director on multiple  successful  films in  the past. Despite  being critically acclaimed, the film grossed only 8 million USD worldwide  in  the  box  office;  which  is,  by  no  means,  a commercially successful movie. So there are other important factors at play here such as the runtime of the movie, the genre, the script, the time of release, etc. which helps a movie become financially successful. Researchers have been working in this domain for a long time to figure out what actually makes a movie financially successful in the Box-Office. But the lack of a large, up-to-date dataset makes their efforts harder. During our literature review we have found numerous works with a very limited number of movies (less than 500). In this work, we present a up-to-date dataset containing detail information of more than 200K movies for researchers and data-analysts to work with. 
    
    

    The contributions of this study are mentioned below.
    \begin{enumerate}
        \item We introduce a larger, richer, and up-to-date movie dataset.
        \item We performed statistical analysis to find out the major factors that influence the amount of revenue generated by the movies.
        \item We implemented and used various machine learning algorithms to predict the revenue generated by a movie. Our prediction models utilize the data that is available during the planning phase of the movie production. So based on our work, producers can make strategic choices in budgeting, selecting the time of release, hiring stars, directors, etc to maximize the chances of success.
    \end{enumerate}

\section{Previous Works} 
Research work for predicting the success of a movie before its release had begun in the early 1980s. Several researchers have tried to determine the parameters that help a movie to become successful at the box office.
\cite{ravid1999information} has claimed that MPAA ratings influence the success of a movie.  According to the authors, movies tend to generate more revenue if they are more \textit{family-friendly}. \cite{kim2013user} found that public opinion is a huge factor for a movie to become successful in the box office. They collected public opinion from the trailer of the movies released on different social media platforms. \cite{choudhery2017social} take into account the tweets and the sentiments behind those tweets from Twitter to make a system that uses polynomial regression to predict the revenue of a movie and claimed that the model gives the best result when the degree of polynomial regression is 6. \cite{shim2017predicting} also considered Twitter data while predicting the opening weekend revenue of any movie using a linear regression model. They worked with the tweets of 67 movies and claimed that the model is 65\% accurate in predicting the gross per day revenue of the movies during the opening week. Some hybrid approaches are also used by some researchers while predicting the gross revenue of any movie [ \cite{dhir2018movie}, \cite{rafipredicting}].
Numerous research has been done on the effect of \textit{Star Power} on movie sales. However, the findings are contradictory. \cite{ravid1999information}, concludes that Star Power plays no role in securing the financial success of a movie. However, \cite{article_1} performs statistical analysis on revenue distribution and performs a KS test, concluding that movies with stars dominate movies without stars when box-office gross is concerned. But, this observation is \textit{Stochastic}, meaning it can be analyzed statistically but cannot be precisely predicted. \cite{selvaretnam2015factors} used two different approaches to calculate the star power: (1) number of nominations/wins of Academy awards by the key players, (2) average lifetime gross revenue of films involving the key players preceding the sample year. They performed regression analysis and found the first approach to be statistically insignificant and the second approach statistically significant. \cite{elberse2007power} conducted a study on the data on Hollywood Stock Exchange and compared movie stock prices before and after the cast was announced. The study found that stars positively influence a movies' stock value. 
Though relatively overlooked compared to star actors, some studies have tried to link between the financial success of a movie with the presence of star directors[\cite{lutter2014creative}, \cite{boccardelli2008critical}, \cite{meiseberg2008we}]. And like with star-power, the findings of these studies are also highly contradictory. \cite{lutter2014creative}, \cite{jung2010does} claimed that top directors along with top actors are associated with the higher probability of financial success of a movie. However, \cite{boccardelli2008critical} and \cite{meiseberg2008we} could not find any relation between a movie's financial success and the presence of a star director.
One of the possible reasons why researchers reach contradictory conclusions on the effect of star power on a movie's financial success might be the fact that researchers use different metrics and approaches to calculate the star power. Researchers such as \cite{simonoff2000predicting} have predicted the total grosses of a movie based on the number of awards the casts of the movie has previously won and reached one result.  Others take into account a stars' or directors' previous financial successes and reach a different result. In this study, we take a hybrid approach to calculate the star and director power of a movie. We consider the number of credits under a stars' or director's name before the sample year, previous financial successes, the average rating of previous credits, number of raters that have given an IMDb rating etc. to calculate the star and director power. And, we found no such previous work that took into account the contributions of story writers, screenplay writers in the financial success of a movie. In this work, we have found their contributions as well to make a movie financially successful to be statistically significant and take into account their contributions when we calculate the star power of a movie.
\section{Dataset}
\subsection{\textbf{Data Collection}}
IMDb (Internet Movie Database) is a website containing information related to \textit{Movies, TV Series, Creative Work, and Video Games}. For this section, we will use the generic term \textit{item} to indicate all these categories. For scraping data from \href{https://www.imdb.com}{IMDb}, we implemented two individual bots. The first bot is for preparing the \textit{summary} data. When the bot is run it goes to a link\footnote{\href{https://www.imdb.com/search/title/?adult=include&count=250}{\textit{\textcolor{blue}{www.imdb.com/search/title/?adult=include\&count=250}}}} containing the top 250 \textit{items} in IMDb sorted by popularity. 
The bot scrapes all the items and accompanying information listed on the page. The scraped attributes per \textit{item} are: \textit{link to the item page, title, year of release, IMDb rating, meta score, certificate, runtime, genre, a brief plot, name of directors, 4 top-billed stars, voters, and gross revenue}. {\color{black}(Please check \textbf{\color{BlueViolet}Figure \ref{fig: summary_structure}} in the Supplementary Figures section for the structure of summary data per item)}. Then the bot goes to the next page by clicking the "\textcolor{blue}{Next »}" link at the bottom of the list, scrapes the next 250 items, and so on. The scraper was run on $18^{th}$ March at 6:00 pm (Bangladesh local time). It took about 8 days for the scrapper to scrape the summary information of all the items listed in IMDb. Our dataset contains summary information of more than 7.5 million (7644784) items listed on the site. 

Our second bot uses the \textit{link} attribute of the item in the \textit{summary} data to go to the IMDb page dedicated to that item and scrapes revenue, budget, names of all the top-billed casts listed on the item page and other relevant information from that page. As the list of information scraped per item page by our second bot is long, we do not mention them here. {Please check \textbf{\color{BlueViolet}Figure \ref{fig: movie_structure}} in the Supplementary Figures section to know all the properties and how they are structured}. We scraped information on more than 200000 movies using this bot. This data was stored in JSON format. The structure of the data elements is depicted on Figure \ref{fig: movie_structure} and \ref{fig: summary_structure}. We make our dataset public so that anyone can work with it. We also open source the code base to load the dataset and dump it into a local MondoDB server. Furthermore, we open-source our scrapper bots so that anyone working in this domain can modify those bots as per their requirements and create their own up-to-date datasets.  
\\[.5cm]
\textit{\textbf{How our dataset is different from the “IMDb 5000 Movie Dataset”}}: The \href{https://www.kaggle.com/carolzhangdc/imdb-5000-movie-dataset}{IMDb 5000 Movie dataset} is one of the most popular datasets among data scientists working in box office analysis and it has been available for a while. So, the question might arise why we needed to create a new dataset from scratch and how our dataset is different from the popular IMDb 5000 dataset. 

\begin{itemize}
    \item Our dataset is much bigger, containing the \textit{summary} data of 7.5 million items and \textit{detail} data of more than 200K movies.
    \item The IMDb 5000 dataset contains information on only 3 top-billed casts. Our dataset has information on all the casts listed on the IMDb page dedicated for a movie.
\end{itemize}

So, our dataset is both \textit{bigger} and \textit{richer}. The only attribute that IMDb 5000 dataset has and our dataset lacks is the \textit{number of Facebook page likes} of the movie directors and 3 stars' verified pages. These numbers of Facebook page likes have been used in calculating the \textit{Star Power} of movies in many works. 
We did not collect these information for 2 reasons.
\begin{enumerate}
    \item Many stars including Jennifer Lawrence, Scarlett Johansson do not have verified Facebook pages. 
    \item It is misguided to account for the current FB likes of a star while analyzing a movie that was released 5 years ago. To calculate the star power of a movie that was released 5 years ago while considering the FB likes, we should consider the number of FB likes of each star in the cast 5 years ago (before the release of the movie). But such data is not available anywhere and will be very hard to collect.   
\end{enumerate}

\section{Data Cleaning and Primary Pre-processing}
\begin{figure}
\centering
    \includegraphics[totalheight=6cm]{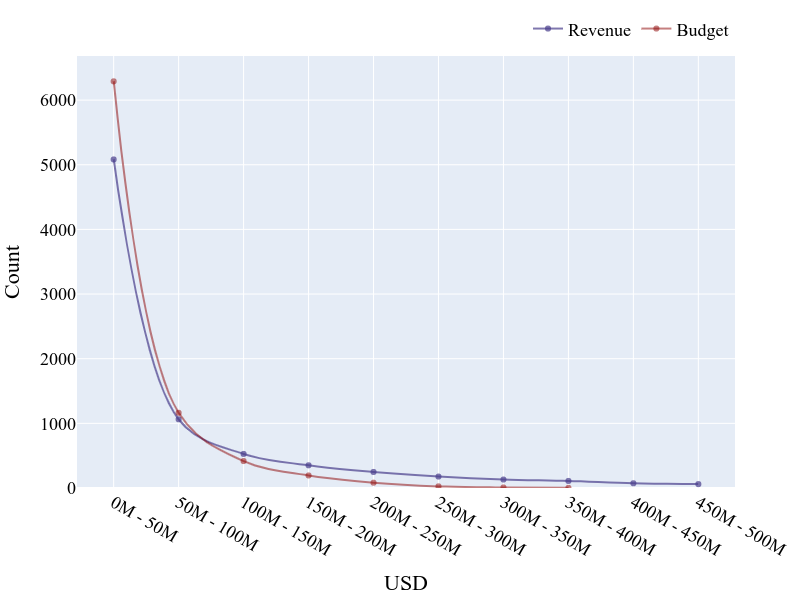}
    \caption{Budget and Revenue distribution of our \textit{cleaned} data. Distribution in 50 million bucket sizes up to 500 million. The values become too small after that.}
    \label{fig: rev_bud_skew}
\end{figure}

We consider the following attributes for a movie.
\begin{itemize}
    \item Revenue (World-wide collection, in USD)
    \item Budget
    \item Runtime (in minute)
    \item Release month
    \item Content Rating (MPAA)
    \item Genre 
    \item Cast 
    \item Directors
    \item Creators
    \item IMDb rating (Not a pre-released feature. But we analyze its association with Revenue to decide whether to include this property when calculating Genre power and Star power. More on that later)
\end{itemize}
We considered all the movies released in the last 4 decades. As we are considering such a long period, we need to adjust the \textit{Revenue} and \textit{Budget} amounts for inflation. We use \href{https://github.com/datadesk/cpi}{\textit{cpi}}, a python library to adjust U.S. Dollars for inflation using the Consumer Price Index (CPI).

$$ Inflation Rate = (\dfrac{CPI_{2}-CPI_{1}}{CPI_{1}}) * 100  $$
where:
\begin{itemize}
     \item CPI\textsubscript{2} - is the CPI in the second period 
     \item CPI\textsubscript{1} - is the CPI in the previous period
\end{itemize}
We removed any movie that did not have \textbf{all} of the mentioned attributes. After cleaning we were left with a total of 8181 movies. We did not perform any data imputations because many attributes (such as Revenue, Budget) have a Lévy distribution with a heavy upper trail and some outliers in the far right corner [\textbf{\color{BlueViolet}Figure \ref{fig: rev_bud_skew}}]. After considering for inflation the mean value of revenue in our cleaned data is 100747045, which is 75.23 percentile of the distribution. This resulted in a heavy rightward skew in the distribution. The budget value is less skewed compared to revenue (mean: 36986670, 67.2 percentile of the distribution). 
This skewed distribution means there is no natural central value to which these distributions converge.

\section{Investigating the Association between Selected Attributes and Revenue} 
In this section, we perform different statistical analysis to check if there exists an association between each of our selected attributes with the movie revenue.

\subsection{\textbf{Budget vs Revenue}} 
Big budget movies are able to incorporate more resources on making the movie compared to lower budget movies. So, we hypothesize, \textit{there will be a positive correlation between movie budget and revenue}. To find out whether budget and revenue are correlated and to which extent, we perform some statistical tests.

First, we normalize Budget values and Revenue values using \textit{Min-Max} scaling. The formula is given below.
$$ X_{norm} = \dfrac{X - X_{min}}{X_{max} - X_{min}} $$

Then we use Spearman Rank Correlation (\text{Spearman rho}) to find the strength of the monotonic relationship between the budget and revenue. We use the \textit{scipy} library of \textit{python}. The test returns the following values.
$$ SpearmanrResult(correlation=0.74, pvalue=0.0) $$
This result says there is a high positive correlation between budget and revenue and the relationship is statistically significant (small p-value). This result supports our hypothesis.

\begin{figure}
\centering
    \includegraphics[totalheight=6cm]{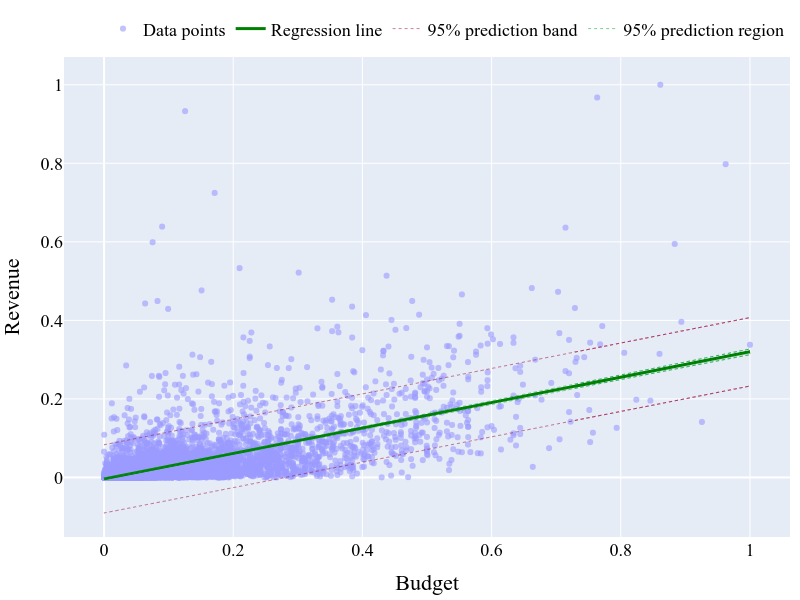}
    \caption{Budget vs Revenue regression \\ \textit{slope=0.32, intercept=-0.0036}}
    \label{fig: bud_rev}
\end{figure}

We further confirm our hypothesis by applying and performing an Ordinary Least Squares (OLS) linear regression on our dataset [\textbf{\color{BlueViolet}Figure \ref{fig: bud_rev}}]. The positive slope of the regression line also aligns with our hypothesis.

\subsection{\textbf{Movie Runtime vs Revenue}} 
Though movies generate revenue from many other sources like selling television rights and streaming rights, advertisements, etc; the bulk of the revenue comes from domestic and foreign ticket sales. And we believe that the duration of a movie affects these ticket sales. Very few people will be interested to go to a movie theater just to watch a short movie with a runtime of 20 minutes. And in this modern era of mobility and the internet, not many people will be willing to sit in a movie theater watching a 4-hour long movie. So we hypothesize, there is a certain threshold until which we will observe a positive correlation between the runtime and revenue. But after that, the generated revenue will start falling. 

To test our hypothesis we perform a Spearman rho test. The test returns a weak correlation value (0.33), but that was expected as our hypothesis suggests that the relationship between movie duration and generated revenue is not \textit{monotonic}.
$$ SpearmanrResult(correlation=0.33, pvalue=1.56e^{-201}) $$

\begin{figure}
\centering
    \includegraphics[totalheight=6cm]{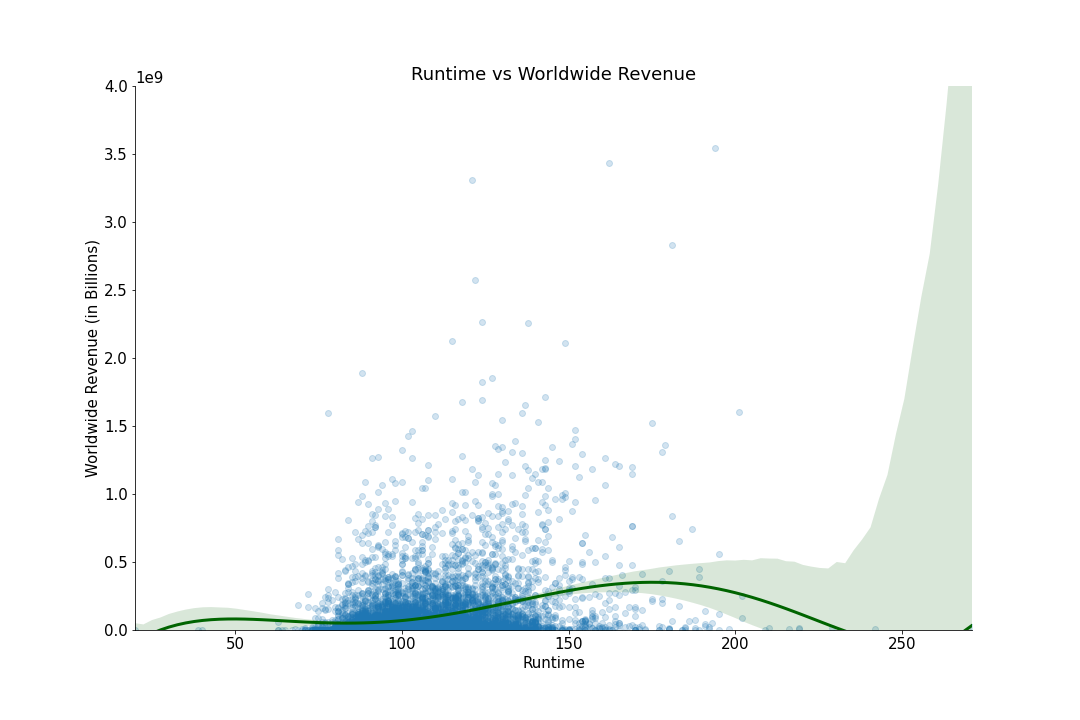}
    \caption{Runtime vs Revenue regression curve}
    \label{fig: runtime_vs_revenue}
\end{figure}

We fitted a regression curve with a six-degree polynomial equation. The data points and the function line is plotted in \textbf{\color{BlueViolet}Figure \ref{fig: runtime_vs_revenue}}. The wavy pattern of the curve, initial rise, and then fall is in line with our hypothesis. From \textbf{\color{BlueViolet}Figure \ref{fig: runtime_vs_revenue}} it is apparent that movies with too short runtime (less than 90 minutes) or too long runtime (more than 200 minutes) usually generate lower revenue.

\subsection{\textbf{Release Month vs Revenue}} 

\begin{figure*}[t!]
    \centering
    \begin{subfigure}[t]{0.5\textwidth}
        \centering
        \includegraphics[height=2.0in]{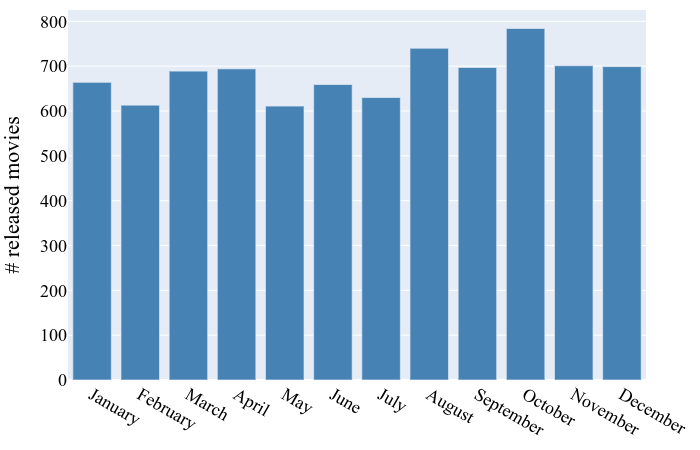}
        \caption{Month-wise number of released movies (1967-2020)}
        \label{fig: month_num}
    \end{subfigure}%
    \begin{subfigure}[t]{0.5\textwidth}
        \centering
        \includegraphics[height=2.0in]{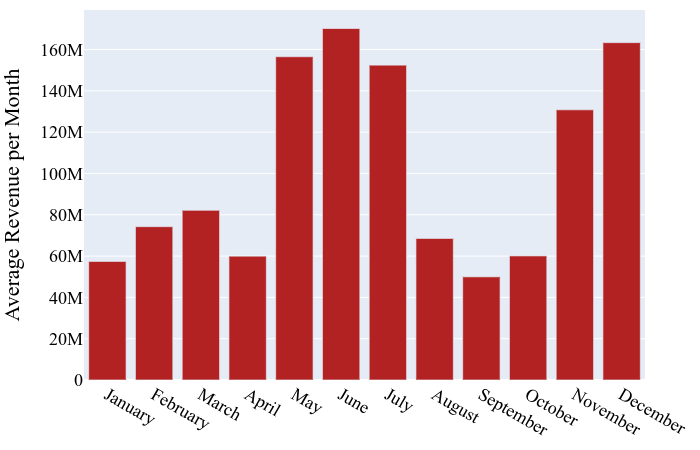}
        \caption{Month-wise average revenue per movie (1967-2020)}
        \label{fig: month_revenue}
    \end{subfigure}
    \caption{}
    \label{fig: month}
\end{figure*}

From \textbf{\color{BlueViolet}Figure \ref{fig: month_revenue}}, it can be easily observed that the movies released in May-July and November-December do significantly better on average compared to movies released during the other months. There might be several reasons for that.
\begin{itemize}
    \item The spike during May-July overlaps with the summer break of schools in USA and Canada. Young audiences in those countries have plenty of free time during these months to go to movie theaters. And the other spike at the end of the year overlaps with the Christmas and holiday season. 

    \item The two \textit{dump} periods according to \href{https://en.wikipedia.org/wiki/Dump_months}{Wikipedia} are January-February and August-September. In January-February, the presence in movie theaters is negatively affected by the award calendar of OSCAR and golden globe, harsh weather, etc. The movie theater presence during August-September is adversely affected because moviegoers under the age of 24 (41\% of the audience) are busy preparing to go back to school.
\end{itemize}

But interestingly the number of movies released during all the months are almost the same. Since the spike times generate more revenue on average, these times are targeted by major production houses with big-budget movies. For example, almost all of the movies of Marvel Cinematic Universe were released during (or shortly before/after) the spike times (except for \textit{Captain America: The Winter Soldier, Black Panther, and Captain Marvel}). So it is possible that other small budget and less ambitious projects target the rest of the months to avoid clash with the big-budget movies.

\subsection{\textbf{Content Rating vs Revenue}} 
MPAA rating is used in the USA to rate a movie's suitability to a certain target audience [\href{https://en.wikipedia.org/wiki/Motion_Picture_Association_film_rating_system}{wiki}]. The ratings are as follows.
\begin{itemize}
    \item G: General Audiences, appropriate for all age groups
    \item PG: Parental Guidance Advised
    \item PG-13: Parents Strongly Cautioned 
    \item R: Restricted
    \item NC-17: Adults only
\end{itemize}
This property was named as \textit{contentRating} in the metadata we scraped from IMDb. We keep the name \textit{Content Rating} as we found 18 different types of such ratings in IMDb items which had \textit{'@type': 'Movie'} in its metadata. 

We choose two \textit{Content Rating}s with the highest number of movies PG-13 and R, and build two separate arrays with revenues of the movies that belonged to each rating. And then we perform a KS test (Kolmogorov–Smirnov test) to check for the equality of the distributions between movies belonging to these two ratings. We used \textit{ks\_2samp} of \textit{python}s' \textit{scipy} library. The test returned the following values
$$ KStestResult(statistic=0.25, pvalue=8.38e^{-98}) $$
The small p-value indicates that the movies' revenue values for these two classes are differently distributed.

\begin{table}
    \begin{center}
        \begin{tabular}{ |c|c|c| } 
             \hline
             Content Rating & \# of Movies\\ \hline
             PG-13      & 2154\\ \hline
             R          & 3753\\ \hline
             PG         & 1245\\ \hline
             TV-MA      & 52\\ \hline
             G          & 184\\ \hline
             Unrated    & 72\\ \hline
             NC-17      & 14\\ \hline
             Not Rated  & 628\\ \hline
             Approved   & 9\\ \hline
             X          & 1\\ \hline
             M          & 2\\ \hline
             M/PG       & 1\\ \hline
             GP         & 5\\ \hline
             TV-PG      & 15\\ \hline
             TV-14      & 36\\ \hline
             TV-Y7      & 3\\ \hline
             Passed     & 1\\ \hline
             TV-G       & 6\\ 
             \hline
        \end{tabular}
    \end{center}
    \caption{Number of movies for each content rating}
    \label{tab: before_clustering}
\end{table}

\begin{figure*}[t!]
    \centering
    \begin{subfigure}[t]{0.5\textwidth}
        \centering
        \includegraphics[height=2.3in]{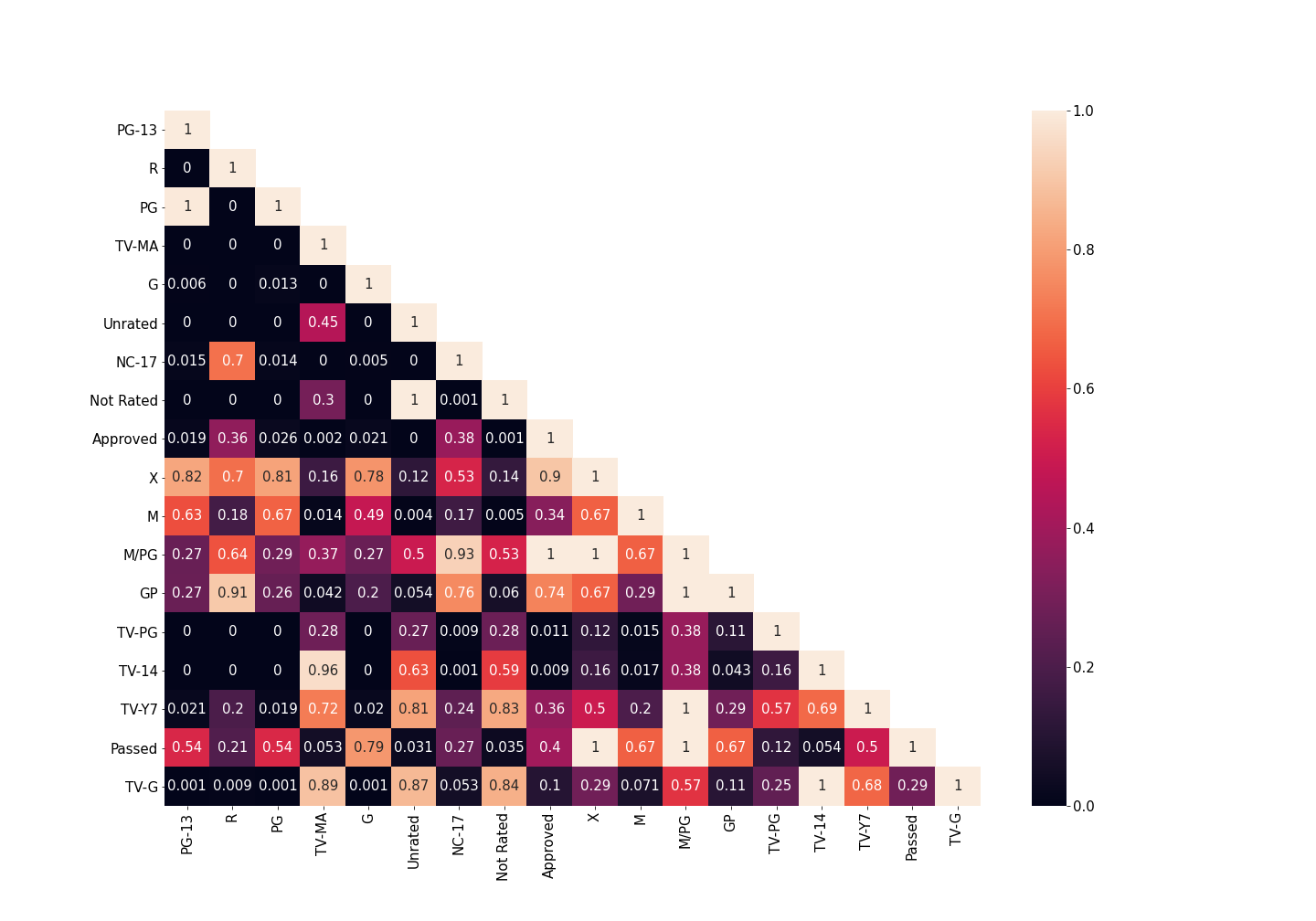}
        \caption{Before Clustering (18 ratings)}
        \label{fig: before_cluster_content}
    \end{subfigure}%
    \begin{subfigure}[t]{0.5\textwidth}
        \centering
        \includegraphics[height=2.3in]{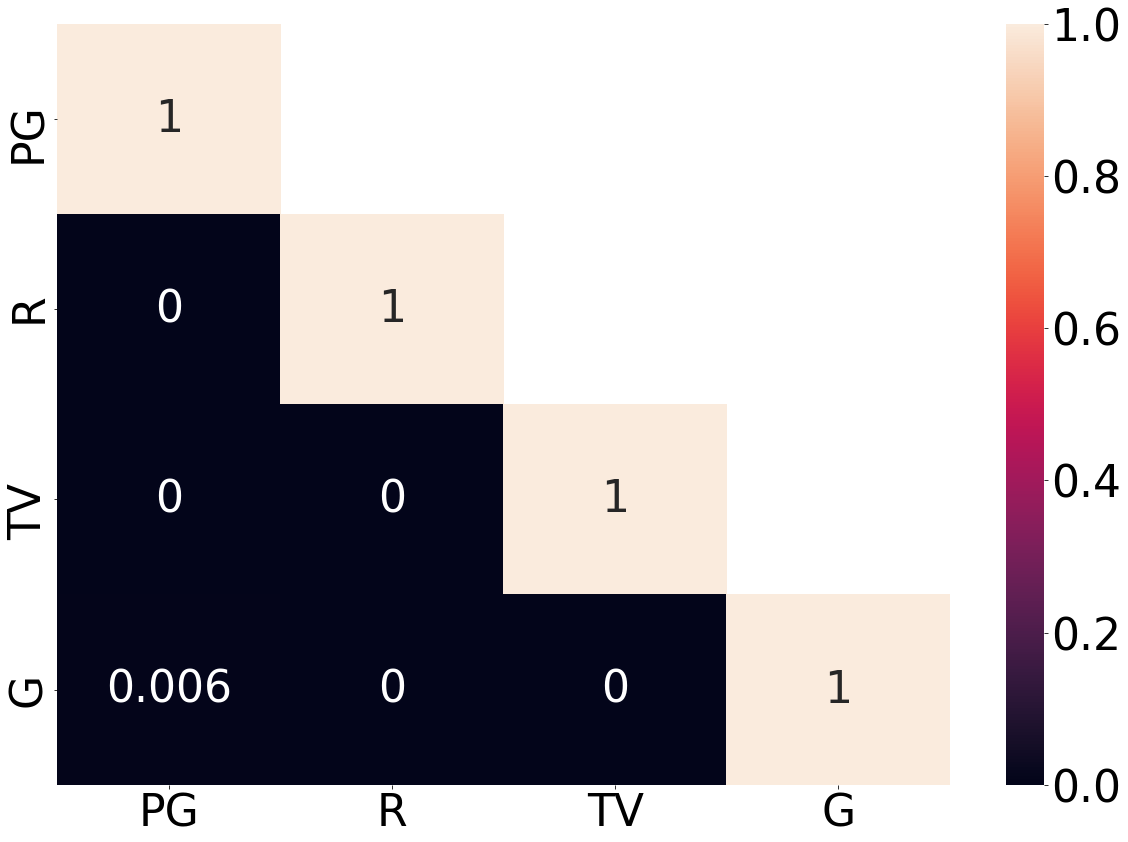}
        \caption{After clustering (6 ratings)}
        \label{fig: after_cluster_content}
    \end{subfigure}
    \caption{Inter class p-value heatmatrix from KS tests performed on pair-wise content rating distributions with the null hypothesis that \textit{different content ratings do not effect the generated revenue distribution}}
\end{figure*}

\textbf{\color{BlueViolet}Table \ref{tab: before_clustering}} shows the distribution of movies in all the 18 \textit{content ratings}. It is apparent that some of those ratings are rare (like X, M etc). So, we perform KS tests for all pairs of ratings [\textbf{\color{BlueViolet}Figure \ref{fig: before_cluster_content}}]. Based on high p-values and low number of movies in a rating class, we make some clusters of such ratings [\textbf{\color{BlueViolet}Table \ref{tab: clustered_content}}]. The inter-class p-value matrix after applying clustering is depicted in \textbf{\color{BlueViolet}Figure \ref{fig: after_cluster_content}}.

\begin{table}
    \begin{center}
        \begin{tabular}{ |c|c| } 
             \hline
             Cluster Label  & Cluster Members\\ \hline
             PG             & PG-13, PG\\ \hline
             R              & R, NC-17, Approved, X, M\\ 
                            & M/PG, GP, Passed, Passed\\ \hline
             TV             & TV-MA, TV-PG, TV-14, TV-Y7,\\ 
                            & TV-G, Unrated, Not Rated\\ \hline
             G               & G \\ \hline
        \end{tabular}
    \end{center}
    \caption{Content Rating Clusters}
    \label{tab: clustered_content}
\end{table}

\subsection{\textbf{Genre vs Revenue}} 
\begin{figure}[h!]
\centering
    \includegraphics[totalheight=6cm]{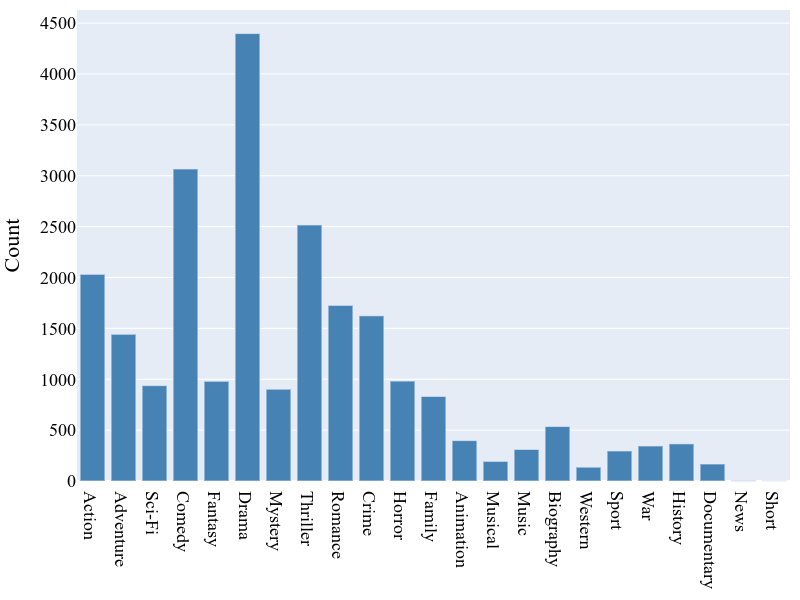}
    \caption{Count of movies by genre.\\ \textit{Note: Same movies may be counted in multiple genres}}
    \label{fig: genre_count}
\end{figure}

IMDb lists multiple genres for some movies, where applicable. For example, the movie \href{https://www.imdb.com/title/tt0371746/}{Iron Man (2008)} has \textit{Action}, \textit{Adventure}, and \textit{Sci-Fi} listed as its genre. In our dataset we found 23 different genres. The count of movies in each genre is depicted in \textbf{\color{BlueViolet}Figure \ref{fig: genre_count}}.

\begin{figure*}[t!]
    \centering
    \begin{subfigure}[t]{0.5\textwidth}
        \centering
        \includegraphics[height=2.3in]{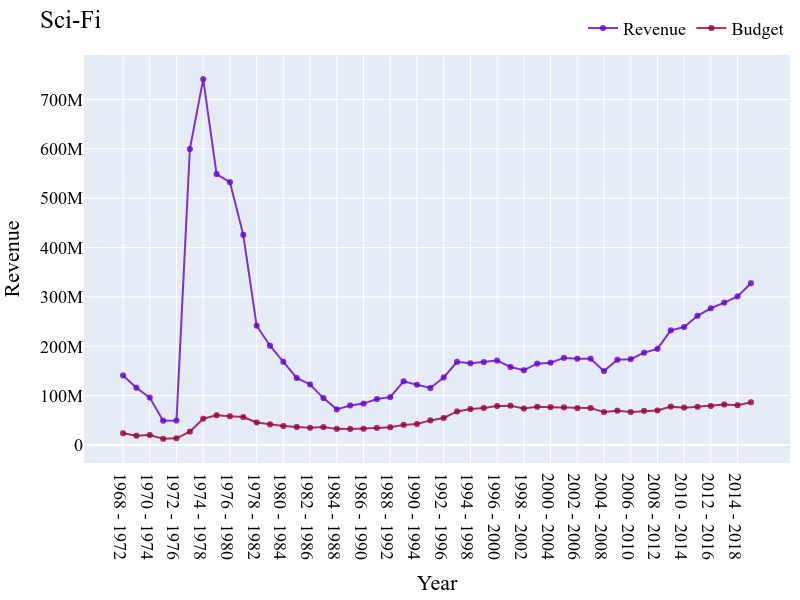}
        \caption{Sci-Fi is gaining popularity in recent years}
        \label{fig: genre_scifi}
    \end{subfigure}%
    \begin{subfigure}[t]{0.5\textwidth}
        \centering
        \includegraphics[height=2.3in]{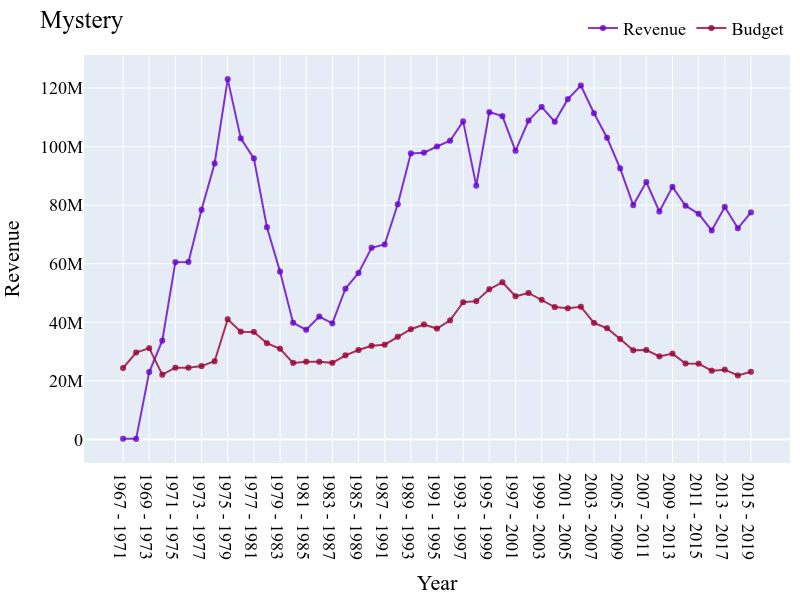}
        \caption{Mystery is losing popularity}
        \label{fig: genre_mystery}
    \end{subfigure}
    \caption{Five year rolling average of the revenue and budget ot Sci-Fi and Mystery genre}
    \label{fig: genre_popularity}
\end{figure*}

We plot a 5-year rolling average of revenues for each genre and notice significant ups and downs. It indicates the increase and decrease in genre popularity with the change of audiences' taste. In \textbf{\color{BlueViolet}Figure \ref{fig: genre_popularity}} we show the graph for Sci-Fi [\textbf{\color{BlueViolet}Figure \ref{fig: genre_scifi}}] and Mystery genre [\textbf{\color{BlueViolet}Figure \ref{fig: genre_mystery}}].

We choose two genres: \textit{Sci-Fi} and \textit{Family} such that a movie has a low probability of having both of these genres. Only 120 movies out of 8181 movies in our cleaned dataset had both Sci-Fi and Family as genres simultaneously. And we perform a KS test to reject the null hypothesis that genre does not have any effect on generated revenues. 
$$KStestResult(statistic=0.102, pvalue=0.0002)$$

{\color{black}The pair-wise KS statistic heatmap is added in the supplementary \textbf{\color{BlueViolet}Figure \ref{fig: genre_ks}}}

\subsection{\textbf{Effect of Star Actors and Renowned Directors/Creators/Production Houses on Revenue}} 
\begin{center}
\vspace{2em}
\textit{
“'A guy stranded on an island' without Tom Hanks is not a movie. With another actor, (the movie ‘Cast Away’) would gross \$40 million. With Tom Hanks it grossed \$200 million. There's no way to replace that kind of star power.”}
- Bill Mechanic, Former Chairman, Twentieth Century Fox
\end{center}

However, the research community has not always been so confident as Mr. Mechanic about the effect of star power on movie revenue generation. Multiple studies have found no link between star power and revenue generation claiming that \textit{"the real star is the movie"}[\cite{article_1}, \cite{ravid1999information}]. And many studies have found a significant positive correlation between the star power and the gross collection of a movie[\cite{selvaretnam2015factors}].

To check if star powers have any effect on the generated revenue, we check the revenue distributions of movies that have at least one star actor and movies that have no star actors. We consider a movie to have a star actor if any of the cast members in that movie had one of the following properties before the year the movie was released.
\begin{itemize}
    \item Has been part of at least one movie that has generated more than 1 Billion in gross revenue (adjusted for inflation)
    \item Has appeared in more than 40 movies.
\end{itemize}
Then we perform a KS test. The results are below.
$$KStestResult(statistic=0.22, pvalue=5.31e^{-82})$$

We also calculate the mean of the revenue distribution of the movies that have a star and do not have a star. The mean revenue of movies that has a star is more than 89 million higher. With that, we can safely reject the null hypothesis and conclude that star power has a significant impact on revenue.

For Director, Creator, and Production we duplicate the experiment we conducted for actor star power. Our findings are depicted in \textbf{\color{BlueViolet}Table \ref{tab: star_power_ks_test}}.

\begin{table*}[t]
    \centering
    \begin{tabular}{|c|c|c|c|c|c|c|}
        \hline
                & Th\textsubscript{\#movies} & Th\textsubscript{revenue} & \#Star Movies & \#No-star Movies & p-value\textsubscript{KS test} & Star\textsubscript{mean} $-$ No-star\textsubscript{mean}\\ \hline
        Actor               &   40  & 1 Billion     & 3230 & 4951 & $5.31e^{-82}$     & 89.3 Million \\ \hline
        Director            &   5   & 100 Million   & 2684 & 5497 & $2.36e^{-246}$    & 138.9 Million \\ \hline
        Creator             &   10  & 400 Million   & 5340 & 2841 & 0.001           & 112.5 Million \\ \hline
        Production Co       &   40  & 1 Billion     & 3337 & 4844 & $2.2e^{-5}$       & 191.2 Million \\ \hline
    \end{tabular}
    \caption{Star Power KS testing thresholds and results\\
    Note: The threshold values $Th_{movies}$ and $Th_{revenue}$ were selected in a trial and error manner until there were had more than 2500 movies in both distributions}
  \label{tab: star_power_ks_test}
\end{table*}

\subsection{\textbf{IMDb rating vs Revenue}}
IMDb allows users of the site to rate a movie out of 1 to 10. However, naturally this feature is not available before the movie is released. We investigated the relationship between this rating and the revenue generated by a movie to check for any significant association. The plan was to incorporate this information in calculating \textit{star power} and \textit{genre} features in case a significant association is found.

The Spearman Rho test returns a significant p-value but a weak correlation. 
$$SpearmanrResult(correlation=0.27, pvalue=3.95e^{-132})$$

\begin{figure}[h!]
\centering
    \includegraphics[totalheight=6cm]{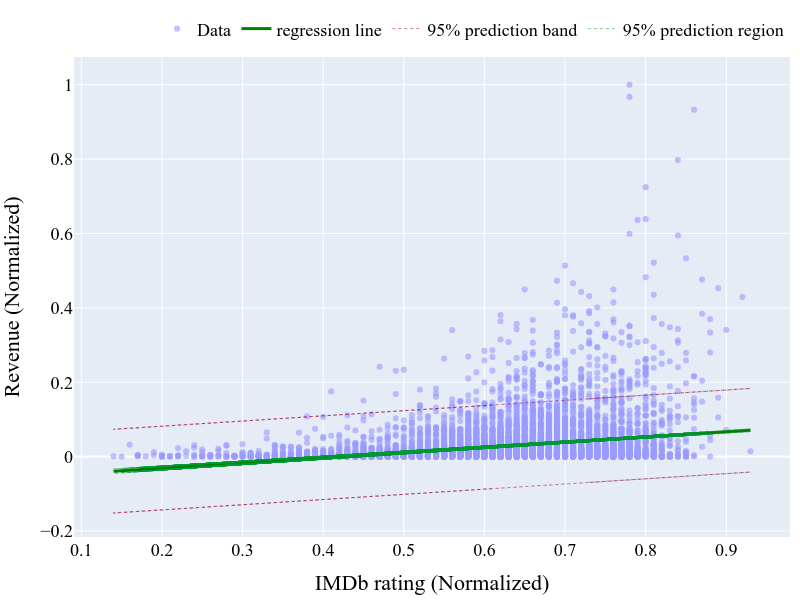}
    \caption{IMDb rating vs Revenue \\ \textit{slope=0.13, intercept=-.06}}
    \label{fig: rating_vs_revenue}
\end{figure}

We apply Min-Max scaling for normalization and find the linear regression line in \textbf{\color{BlueViolet}Figure \ref{fig: rating_vs_revenue}}. From the figure, we can observe that movies with higher revenue usually have a higher rating. But there are a lot of movies with a high rating that has done poorly in generating revenue, hence the gentle slope.

\begin{figure}[h!]
\centering
    \includegraphics[totalheight=6cm]{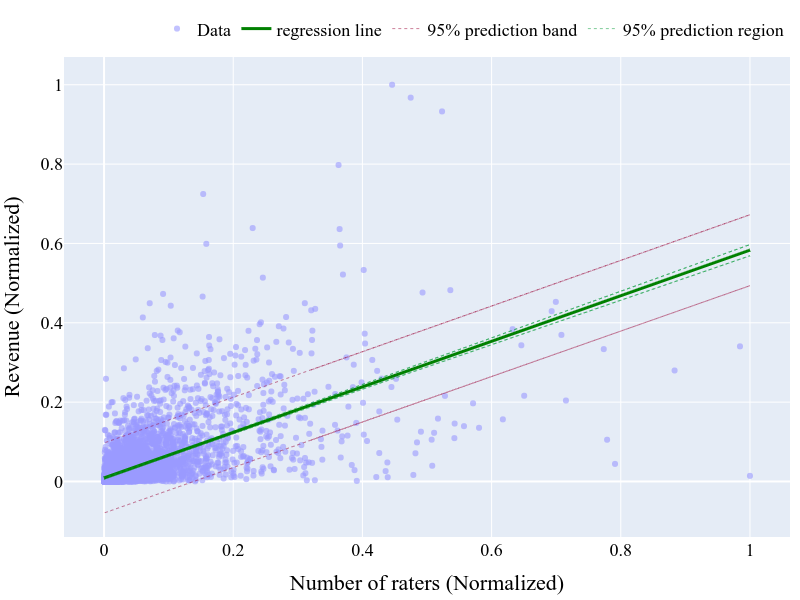}
    \caption{Number of Raters vs Revenue \\ \textit{slope=0.57, intercept=-.01}}
    \label{fig: rater_vs_revenue}
\end{figure}

We also consider the effect of the number of people that have given a rating (raters) on movie revenue. Surprisingly, the Spearman Rho test gives a much higher correlation value for the number of raters vs revenue.
$$SpearmanrResult(correlation=0.78, pvalue=0.0)$$
The slope of linear regression is also much steeper [\textbf{\color{BlueViolet}Figure \ref{fig: rater_vs_revenue}}]. This means that the number of raters is more positively correlated with movie revenue and thus a better indicator of movie revenue prediction. So, we considered both IMDb rating and the number of raters when we calculated Star power and Genre effect.

To summarise, the attributes that have numeric values (such as; budget, runtime) we mainly performed linear/polynomial regression to check for correlation. The attributes that do not have meaningful numeric values (such as: content rating, genre etc) we perform Kolmogorov–Smirnov test to check for the equality of the distributions and use the test result to reject the null hypothesis if we could. The summary of our findings are listed below.

\begin{itemize}
    \item Generated revenue is positively correlated with the budget of the movie.
    \item The correlation with movie runtime is non-linear. From our analysis, we observed that movies with too short runtime (less than 90 minutes) or too long runtime (more than 200 minutes) usually generate lower revenue. 
    \item Movies released in season May-July and November-December tend to better in Box-Office compared to movies released in other months.
    \item Low p-value in KS tests indicate that content rating and genre has an impact on generated revenue.
    \item We also took a number of movies from our dataset and distributed the sample into 2 classes; movies that have star cast and movies that do not. We found that in our sample the movies with star actors and no-star movies had a mean revenue difference of almost 90 million USD. We also analyze the effect of IMDb rating and the number of raters on the generated revenue. In both of the cases we have found positive associations. But, we have observed a stronger correlation of generated revenue with number of raters compared to IMDb raing.
\end{itemize}

In the next sections, we use the insights from our statistical analysis to apply machine learning algorithms to predict the gross revenue. 
\section{Feature Engineering}

\subsection{Feature Extraction} 
In this section, we discuss how we constructed our feature vector for training machine learning algorithms. 

\subsubsection{Budget} We take the budget in USD. Like revenue, the  budget amount was adjusted for inflation using Consumer Price Index (CPI).

\subsubsection{Month} For month, we take 2 values: an integer number indicating the month order and the average revenue generated by movies released on that month.

\subsubsection{Runtime} An integer number indicating the runtime of the movie in minutes.

\subsubsection{Content Rating} After applying clustering on content rating we had 4 different clusters [\textbf{\color{BlueViolet}Table \ref{tab: clustered_content}}]. We encode those values to numbers (PG=0, R=1, Not Rated=2, G=3). For each movie get the cluster number of the movie's content rating and apply one hot encoding. For example, if the content rating of a movie is PG-13, it belongs to the PG cluster (cluster no 0). After applying one hot encoding, we get the vector [1, 0, 0, 0], where the $0^{th}$ index is 1 and values of other indices are zero. So, for each movie, we get a one-dimensional vector of size 4 expressing its content rating.

\subsubsection{Calculating Star Power}
To calculate the star power of a movie, we consider 10 top-billed actors. For each actor, we consider the following properties.
\begin{itemize}
    \item Total number of movies this actor has appeared on before the release year of the movie, for which we are calculating the star power.
    \item Total revenue earned in those movies.
    \item Average revenue earned per movie
    \item Total number of raters rated on IMDb on those movies.
    \item Average number of raters per movie
    \item Average IMDb rating of those movies.
\end{itemize}
So, for each actor, we get 6 feature values. 60 feature values for 10 actors. In case IMDb does not list 10 actors for a movie, we apply zero padding to get 60 feature values for actors.

For directors, creators, and production companies our approach to calculate star power is almost the same. We consider 3 directors (18 feature values) and 5 for both creator and production companies (30 feature values each). Like actors, in case IMDb does not list up to 3 directors (or 5 creators/production companies) we apply zero padding so that all the movies have the feature vectors of same dimensions.

\subsubsection{Calculating Actor Familiarity}
In the previous section, we calculated star power features for a movie based on individual seniority and previous successes. But this star power value does not account for team dynamics and characteristics. A good recipe for ensembling a great cast for a successful movie could be to hire actors who have already appeared together in numerous popular and financially successful movies. For example, before the movie \textit{La La Land} (Released in 2016), the actor pair Ryan Gosling and Emma Stone had appeared in two financially successful movies together \textit{Crazy, Stupid, Love}(2011) and \textit{Gangster Squad}(2013); both movies with a gross collection of more than 100 million. And the movie \textit{La La Land} was a blockbuster hit, bagging almost 450 million worldwide with just a 30 million budget. Surely, we cannot attribute this huge success solely to the acting prowess and team chemistry of the star pair, but certainly, it had a role to play. 

Our calculation for the team dynamics of a movie is influenced by the approaches taken by \cite{lash2016early}. For each movie, we build a dynamic collaboration network among the actors of a movie based on their co-appearances in movies released  previously. In this network, each node represents an actor. An undirected edge is drawn between two nodes if the actors represented by the nodes have appeared in at least one movie. The weight of the edge is the number of movies in which the star pair had collaborated. We used this graph to calculate the structural similarity between cast members. We represent this network as an adjacency matrix and calculate the average cosine similarity between each pair of actors in the cast (pair of rows in the matrix). The equation is given below. In the equation, $N$ denotes the number of actors in the cast, $Actor_i \cdot Actor_j$ denotes the dot product between two actors and $||Actor_i||$ denotes the magnitude of the row vector in the adjacency matrix for $Actor_i$.

$$ Familiarity = \frac{1}{\binom{N}{2}} \displaystyle\sum_{i=1}^{N-1}\displaystyle\sum_{j=i+1}^{N} \frac{Actor_i \cdot Actor_j}{||Actor_i||\times||Actor_j||} $$

Furthermore, for every two stars in the movie, we take the average revenue, the average number of IMDb raters and the average IMDb rating of the movies where the star pair has appeared together previously. And we take the maximum of the pair average revenues, IMDb raters and IMDb rating as features. So, we calculate four values to capture the actor familiarity of the cast for a movie.

\subsubsection{Genre} We calculate how movies release with similar genre has performed in recent 5 years. The way this was calculated is almost similar as to how we calculated the star power for actors, directors, and creators. We calculate feature values for up to 5 genres and take the same 6 feature values for each genre. The only difference is we consider the movies released in the last 5 years before the release year of our target movie (for which we are calculating extracting the genre features).
\newline
\newline
In total we have a vector of 180 dimensions for each movie. A summary breakdown of which feature contributes how many values is depicted in \textbf{\color{BlueViolet}Table \ref{tab: feature_dimension}}

\begin{table}
    \begin{center}
        \begin{tabular}{ |c|c| } 
             \hline
             \textbf{Attribute Name}         & \textbf{\# Feature Values}\\ \hline
             Budget(USD)            & 1 \\ \hline
             Runtime(minutes)       & 1 \\ \hline
             Month                  & 2 \\ \hline
             Content Rating         & 4 \\ \hline
             Genre [5]              & 30 \\ \hline
             Actor [10]             & 60 \\ \hline
             Director [3]           & 18 \\ \hline
             Creator [5]            & 30 \\ \hline
             Production Company [5] & 30 \\ \hline
             Actor Familiarity [4]  & 4 \\ \hline
             \hline
             Total                  & 180 \\ \hline
        \end{tabular}
    \end{center}
    \caption{Feature Dimension}
    \label{tab: feature_dimension}
\end{table}

\subsection{Feature Normalization} Our feature values are of different scales. Budget value can be up to 350 Million whereas the IMDb rating scale is 1 to 10. So, we apply Min-Max scaling on each of the attributes so that each value remains within the range 0 to 1. 

\section{Revenue Prediction}

Many studies categorized the movies in only two categories based on their revenues: \textit{Hit} or \textit{Flop}; providing very limited information. In this work, we approach this problem as a multi-class classification problem. We classify movies into 10 different categories ranging from \textit{Flop} to \textit{Blockbuster} based on their generated revenue. 

\subsection{Defining Classes} 
We follow similar approaches taken by \cite{sharda2006predicting} to define our classes [\textbf{\color{BlueViolet}Table \ref{tab: classes}}]. 
\begin{table}
    \begin{center}
        \begin{tabular}{ |c|c|c| } 
             \hline
             \textbf{Class Label} & \textbf{Revenue Range} & \textbf{\# Movies}\\ 
              & (in Millions USD) & \\ \hline
             0      & $<1$                  & 1608\\    \hline
             1      & $\geq1$   and $<10$   & 1432 \\   \hline
             2      & $\geq10$  and $<20$   & 734\\     \hline
             3      & $\geq20$  and $<40$   & 942\\     \hline
             4      & $\geq40$  and $<65$   & 761\\     \hline
             5      & $\geq65$  and $<100$  & 668\\     \hline
             6      & $\geq100$ and $<150$  & 527 \\    \hline
             7      & $\geq150$ and $<225$  & 490 \\    \hline
             8      & $\geq225$ and $<350$  & 417\\     \hline
             9      & $\geq350$ & 602\\ \hline
        \end{tabular}
    \end{center}
    \caption{Class definition and movie distribution in different classes}
    \label{tab: classes}
\end{table}

These classes have an \textbf{ordinal} relationship among themselves. The difference between class 1 and class 5 is greater than the difference between class 4 and class 5. A number of previous works on revenue prediction has tried to do so while ignoring this important property. It might be because \textit{typical} machine learning toolkits/packages usually assume that the classes are unordered for a classification problem and do not have builtin classifiers or algorithms implemented in them to handle Ordinal Classification problems. In this paper, we take two different approaches in solving this Ordinal Classification problem.
 
\subsection{Balancing the classes} It is observed in \textbf{\color{BlueViolet}Table \ref{tab: classes}} that, the distribution of movies in the classes are highly imbalanced. The highest number of movies belong to class 0 (1608) and the lowest number of movies are in class 8 (417). To make our classes balanced, we randomly select 417 movies from each of the classes. So, to train and test our approaches we use a dataset of 4170 movies, equally distributed among the classes.

\subsection{Train-Test Split} We randomly choose 30\% of our data (1251 movies) and separate them for testing. The other 70\% (2919 movies) were used to train different models. We report the models' performance on the test data.

\subsection{Performance Evaluation Matrices} Our problem at its core is a classification problem and we ensure that the classes are evenly balanced. So, we simply report what percentage of test dataset were assigned by our models to appropriate classes. In this work, we report the \textit{bingo} accuracy and \textit{one class away} accuracy for each of the models. The difference between \textit{bingo} and \textit{one class away} accuracy is that in \textit{one class away} we consider a label to be predicted correctly if the predicted class falls within one class distance of the target class.  That means if the correct class of a movie is 5($\geq65$  and $<100$), then even if a model predicts the class as 4 or 6, according to  \textit{one class away} accuracy it will be considered as a correct prediction. Whereas in \textit{bingo} accuracy, any prediction which does not match with the correct prediction, no matter how close to the correct label the prediction is will be considered as an unsuccessful prediction.

\subsection{Approach 1} 
In our first approach, we implemented the method suggested in the paper \textit{A Simple Approach to Ordinal Classification} by \cite{frank2001simple}. Our number of classes are 10. We implement 9 binary classifiers to predict if the target (revenue of the movie) is greater than 9 different values ($0^{th}$ classifier: predict if revenue exceeds the limit of $0^{th}$ class, $1^{st}$ classifier: predict if revenue exceeds the limit of $1^{st}$ class and so on). Then we aggregate them and store them in a python dictionary. We tested this approach by implementing different classifier models from \textit{sklearn} library. We used these classifiers out of the box with little or no hyperparameter tuning. The classifiers we used in this approach and their respective hyperparameters are mentioned below.

\subsubsection{Logistic Regression} Logistic Regressions (or Logit Regressions) are appropriate regression analysis models to use when the target or dependent variable can have only 2 possible types (pass/fail, win/loss etc). We use Logistic Regression classifiers with parameter $max\_iter=150$ to determine whether a movie exceeds the high value of a certain range or not. Logistic Regression classifiers gave us a \textit{bingo} accuracy of 26.9\% and a \textit{one class away} accuracy of 59.3\%.

\subsubsection{SVC} Support Vector Machines (SVM) are a set of supervised learning approaches for solving classification and regression problems. SVMs are highly effective in high dimensional spaces. For this property,  they are heavily used in different classification tasks even though they are comparatively more computationally expensive to train. We use SVC(Support Vector Classification) models shipped in \textit{sklearn.svm} module with no hyperparameter tuning. It gave us a \textit{bingo} accuracy of 23.9\% and a \textit{one class away} accuracy of 58.0\%.

\subsubsection{Random Forests} Random Forest classifiers are a type of \textit{ensemble} classifiers that fit a number of decision trees on different sub-samples of the dataset and uses an averaging method. The goal of these ensemble methods is to build several estimators independently for a specific task and then average their predictions to improve generalizability over a single estimator. We initialize \textit{sklearn}'s \textit{RandomForestClassifier} with 100 decision trees for our task. It gave us a \textit{bingo} accuracy of 27.2\% and a \textit{one class away} accuracy of 56.9\%.

\subsubsection{MLPClassifier} Neural Networks are powerful tools that endeavors to learn the underlying pattern or structures of data using an approach that mimics the operation of the human brain. \textit{MLPClassifier}s rely on neural networks to perform the task of classification. In Multi Layer Perceptron (MLP) Networks the neurons/perceptrons are arranged layer-wise. Usually,  the layers can be of 3 types. An input layer,  followed by a number of hidden layers and finally the output layer. The number of neurons in the input layer is equal to the number of dimensions in the data. In our case, that is 180. And the number of neurons in a binary classifier MLP network is 1; in our case, it will fire if the classifier predicts that the movie revenue will exceed the high range of the revenue bucket [see \textbf{\color{BlueViolet}Table \ref{tab: classes}}] for which the classifier was built. The choice of  the number of hidden layers and the number of neurons in those layers is not so definite, however. But there are various rules of thumbs. \cite{hidden} in his book \href{https://books.google.com.bd/books?id=Swlcw7M4uD8C&printsec=frontcover&dq=Introduction+to+Neural+Networks+for+Java,+Second+Edition&hl=it&sa=X&redir_esc=y#v=onepage&q=Introduction}{\color{black} Introduction to Neural Networks for Java, 2nd Edition} said, an MLP network with 2 hidden layers \textit{can represent an arbitrary decision boundary to arbitrary accuracy with rational activation functions and can approximate any smooth mapping to any accuracy.}[\textbf{\color{BlueViolet}Table 5.1} in the mentioned book].


So, we determine the number of hidden layers in our classifier networks be 2. For the number of neurons in those hidden layers, we follow one of the 3 rules of thumbs mentioned in the same book. We set that number to be 2/3 the size of the input layer, plus the size of the output layer. MLPClassifier fails to give us a satisfactory accuracy (\textit{bingo} accuracy of 21.2\% and a \textit{one class away} accuracy of 50.34\%). 

\subsubsection{KNN} K-Nearest Neighbor(KNN) is a simple and versatile machine learning algorithm that can be used in both classification and regression. KNNs embrace a feature similarity based approach. To assign a class to a data point, KNN queries the K data points in the training set that are nearest to the query data in the feature space and assigns a class based on maximum voting. We initialize KNN with all the default parameters of \textit{sklearn} library. But as KNNs are not suited for handling large dimensional data the KNN model fails to give a satisfactory accuracy (\textit{bingo} accuracy of 18.7\% and a \textit{one class away} accuracy of 46.4\%). 

\subsubsection{Ensemble} 
\begin{figure}[h!]
\centering
    \includegraphics[height=2.3in]{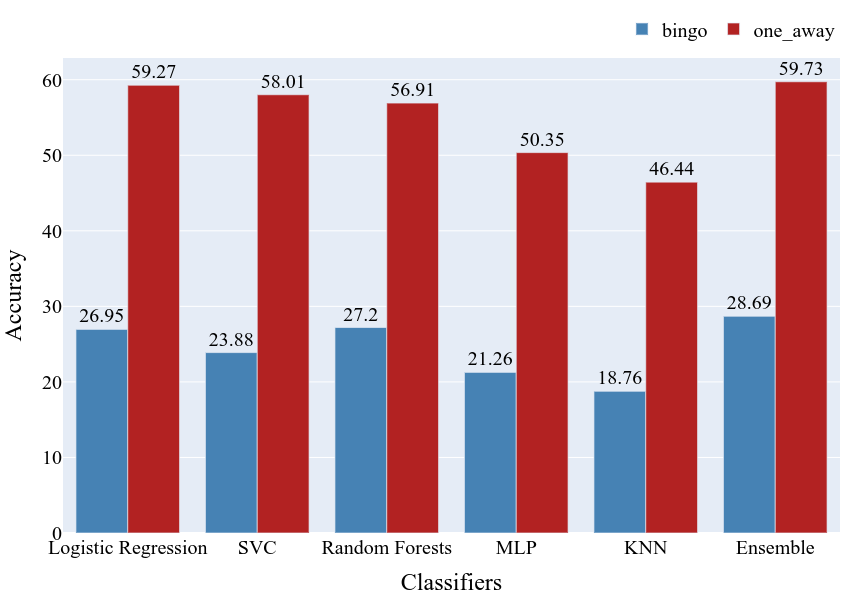}
    \caption{Approach 1 classifiers}
    \label{fig: sklearn}
\end{figure}

We simply aggregate all our classifier models except the KNN model (the only model that gave a one away accuracy rate $< 50\%$) and take the maximum vote as the prediction. In case there is a tie, we take the smallest class. Our ensemble classifier gave us a \textit{bingo} accuracy of 28.7\% and a \textit{one class away} accuracy of 59.7\%
\newline
\newline
The performance comparison of these classifiers are depicted on \textbf{\color{BlueViolet}Figure \ref{fig: sklearn}}

\subsection{Approach 2} 
\begin{figure}[h!]
\centering
    \includegraphics[height=2.3in]{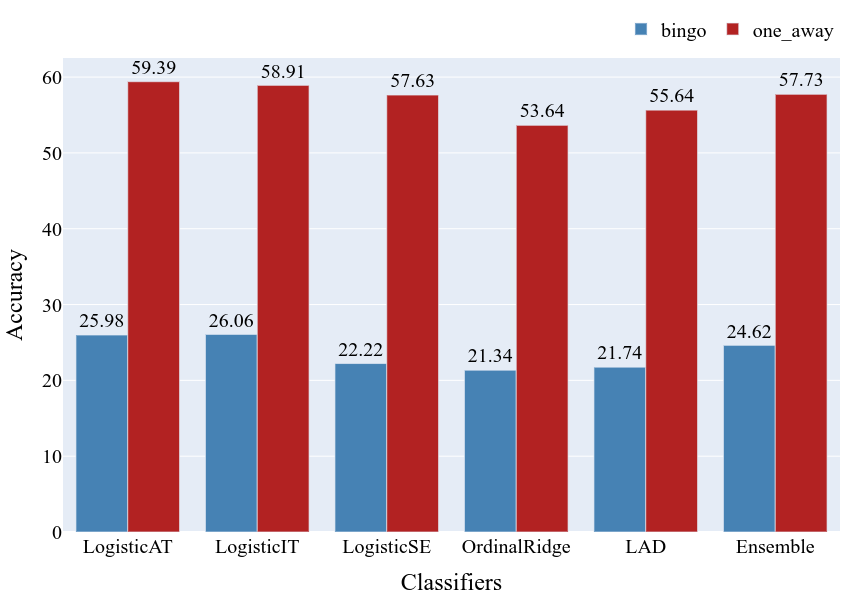}
    \caption{Approach 2 models}
    \label{fig: mord}
\end{figure}

\cite{pedregosa2017consistency} in their paper \textit{On the consistency of ordinal regression methods} presented a python library \textit{mord}, which has different ordinal classification methods implemented in it. We initialize these models with their default hyperparameters and plug into them our train data. The performance of those ordinal classifiers in our dataset are reported in \textbf{\color{BlueViolet}Figure \ref{fig: mord}} 

\vspace*{-.3cm}
\section{Conclusion}
This study introduces a new dataset and tries to statistically analyze which attributes influence generated revenue. Then we tried to use different machine learning algorithms to predict the success of a movie based on the attributes that are available before its release to help movie producers take better strategic decisions. However, the machine learning algorithms did not achieve a satisfactory accuracy (highest bingo accuracy: 28.69\%, and one away accuracy: 59.73\%). This indicates the high level of uncertainty surrounding the prediction of revenue in motion picture industry. For instance, the 1990 film ‘Home Alone’ was a huge success in the box office,  grossing over 476 million USD despite having no renowned names in its cast and having a moderate budget of 18 million USD only. This proves that anything can happen in the motion picture industry. As Jack Valenti, former CEO of the Motion Picture Association of America, once said, \textit{"No one can tell you how a movie is going to do in the marketplace ...not until the film opens in darkened theatre and sparks fly up between the screen and the audience"}. However, in our literature review we have reported one study, \cite{shim2017predicting} have claimed to achieve 65\% accuracy rate in predicting per day movie revenue during the first week. But we would like to point out that they worked with tweets of only 67 movies. Both the approach adnd scope of our work are significantly different compared to their work. Although we could not reach a satisfactory accuracy rate, this study reveals several important findings. We found that the number of people who have given an IMDb rating is more closely associated with revenue, compared to the actual IMDb rating. We used this insight to propose a novel method of calculating star power. There are various scopes for extending this work. Our biggest contribution is the dataset, but we could use only a small fraction of it to analyze statistical relationships and train/test our machine learning algorithms. Data from other sources such as \href{https://www.the-numbers.com/}{www.the-numbers.com}, \href{https://www.boxofficemojo.com/}{www.boxofficemojo.com}, etc. can be incorporated into the dataset to make it more complete. Also, we did not take into account the media buzz and advertising campaigns in social media sites like Facebook and Twitter because of the unavailability of appropriate data. Also, it will be interesting to use Natural Language Processing techniques on youtube comments after the trailer of a movie is released and check for association with the revenue the movie ultimately generates. 

\section{Resource Availability}
\hspace*{-.6cm} \textbf{Dataset:} \href{https://drive.google.com/drive/folders/18uJAC1bEKpHUr5BEG2pi4CgpRXdL2HGa?usp=sharing}{\color{blue}Google Drive}\\
\textbf{Code:}\\
\hspace*{.5cm}\textbf{Data Scraper:} \href{https://github.com/arnab-api/IMDb-Scraper}{\color{blue} github.com/arnab-api/IMDb-Scraper}\\
\hspace*{.5cm}\textbf{Analysis:} \href{https://github.com/arnab-api/Movie-Analysis}{\color{blue} github.com/arnab-api/Movie-Analysis}\\

\section*{Acknowledgement}
We would like to thank the Department of Computer Science and Engineering of Shahjalal University of Science and Technology for lending us server PCs to run our scrapper bots.

\bibliographystyle{unsrt}  
\bibliography{references}  

\begin{thebibliography}{10}

\bibitem{lash2016early}
Michael~T Lash and Kang Zhao.
\newblock Early predictions of movie success: The who, what, and when of
  profitability.
\newblock {\em Journal of Management Information Systems}, 33(3):874--903,
  2016.

\bibitem{ravid1999information}
S~Abraham Ravid.
\newblock Information, blockbusters, and stars: A study of the film industry.
\newblock {\em The Journal of Business}, 72(4):463--492, 1999.

\bibitem{kim2013user}
Daehoon Kim, Daeyong Kim, Eenjun Hwang, and Hong-Gu Choi.
\newblock A user opinion and metadata mining scheme for predicting box office
  performance of movies in the social network environment.
\newblock {\em New review of hypermedia and multimedia}, 19(3-4):259--272,
  2013.

\bibitem{choudhery2017social}
Deepankar Choudhery and Carson~K Leung.
\newblock Social media mining: prediction of box office revenue.
\newblock In {\em Proceedings of the 21st International Database Engineering \&
  Applications Symposium}, pages 20--29, 2017.

\bibitem{shim2017predicting}
Steve Shim and Mohammad Pourhomayoun.
\newblock Predicting movie market revenue using social media data.
\newblock In {\em 2017 IEEE International Conference on Information Reuse and
  Integration (IRI)}, pages 478--484. IEEE, 2017.

\bibitem{dhir2018movie}
Rijul Dhir and Anand Raj.
\newblock Movie success prediction using machine learning algorithms and their
  comparison.
\newblock In {\em 2018 First International Conference on Secure Cyber Computing
  and Communication (ICSCCC)}, pages 385--390. IEEE, 2018.

\bibitem{rafipredicting}
Quazi Ishtiaque~Mahmud, Nuren~Zabin Shuchi, Fazle~Mohammed Tawsif, Asif
  Mohaimen, and Ayesha Tasnim.
\newblock A machine learning approach to predict movie revenue based on
  pre-released movie metadata.
\newblock {\em Journal of Computer Science}, 16(6):749--767, Jun. 2020.

\bibitem{article_1}
Arthur De~Vany and W.~Walls.
\newblock Uncertainty in the movie industry: Does star power reduce the terror
  of the box office?
\newblock {\em Journal of Cultural Economics}, 23:285--318, 11 1999.

\bibitem{selvaretnam2015factors}
Geethanjali Selvaretnam and Jen-Yuan Yang.
\newblock Factors affecting the financial success of motion pictures: what is
  the role of star power?
\newblock 2015.

\bibitem{elberse2007power}
Anita Elberse.
\newblock The power of stars: Do star actors drive the success of movies?
\newblock {\em Journal of marketing}, 71(4):102--120, 2007.

\bibitem{lutter2014creative}
Mark Lutter.
\newblock Creative success and network embeddedness: Explaining critical
  recognition of film directors in hollywood, 1900--2010.
\newblock 2014.

\bibitem{boccardelli2008critical}
Paolo Boccardelli, Federica Brunetta, and Francesca Vicentini.
\newblock What is critical to success in the movie industry? a study on key
  success factors in the italian motion picture industry.
\newblock 2008.

\bibitem{meiseberg2008we}
Brinja Meiseberg, Thomas Ehrmann, and Julian Dormann.
\newblock We don’t need another hero—implications from network structure
  and resource commitment for movie performance.
\newblock {\em Schmalenbach Business Review}, 60(1):74--98, 2008.

\bibitem{jung2010does}
Sang-Chul Jung and Myeong~Hwan Kim.
\newblock Does the star power matter?
\newblock {\em Applied Economics Letters}, 17(11):1037--1041, 2010.

\bibitem{simonoff2000predicting}
Jeffrey~S Simonoff and Ilana~R Sparrow.
\newblock Predicting movie grosses: Winners and losers, blockbusters and
  sleepers.
\newblock {\em Chance}, 13(3):15--24, 2000.

\bibitem{sharda2006predicting}
Ramesh Sharda and Dursun Delen.
\newblock Predicting box-office success of motion pictures with neural
  networks.
\newblock {\em Expert Systems with Applications}, 30(2):243--254, 2006.

\bibitem{frank2001simple}
Eibe Frank and Mark Hall.
\newblock A simple approach to ordinal classification.
\newblock In {\em European Conference on Machine Learning}, pages 145--156.
  Springer, 2001.

\bibitem{hidden}
Jeff Heaton.
\newblock {\em Introduction to Neural Networks for Java}.
\newblock 2005.

\bibitem{pedregosa2017consistency}
Fabian Pedregosa, Francis Bach, and Alexandre Gramfort.
\newblock On the consistency of ordinal regression methods.
\newblock {\em Journal of Machine Learning Research}, 18:1--35, 2017.

\end{thebibliography}

\clearpage

\onecolumn

\section{Supplimentary Figures}

\begin{figure}[hbt!]
\centering
    \includegraphics[height=7in]{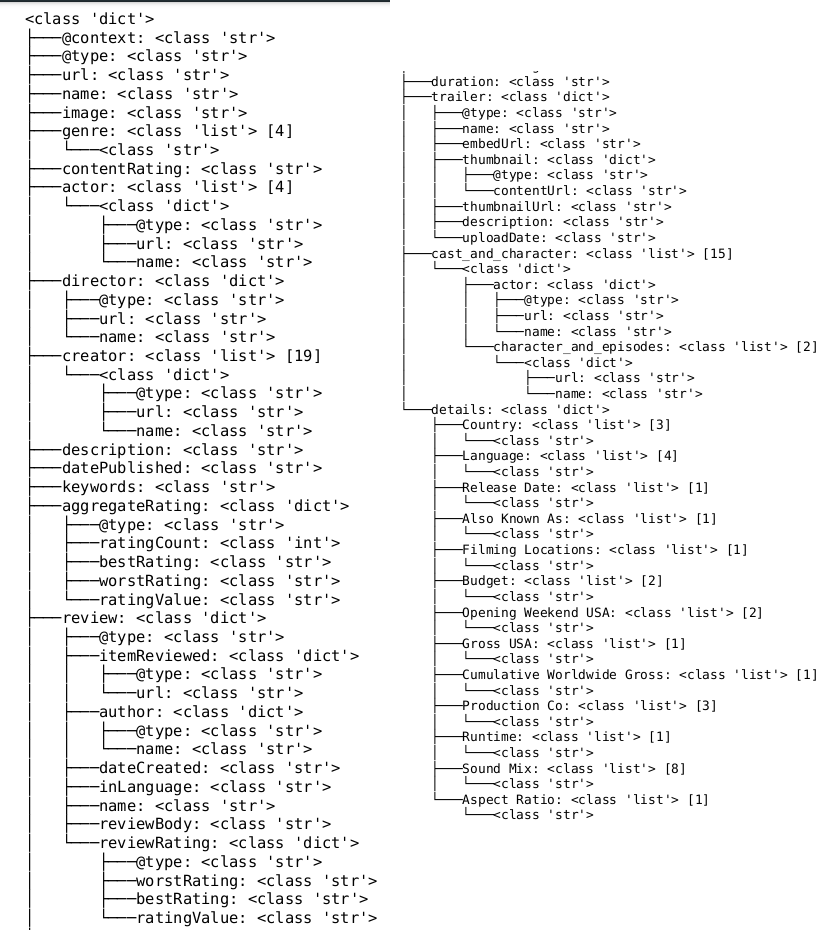}
    \vspace{2em}
    \caption{Structure of movie data}
    \label{fig: movie_structure}
\end{figure}

\begin{figure}[hbt!]
\centering
    \includegraphics[height=3in]{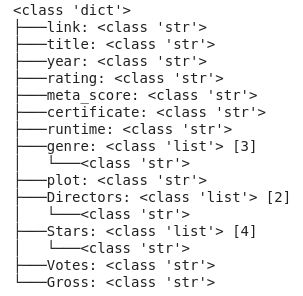}
        \vspace{2em}
    \caption{Structure of summary data}
    \label{fig: summary_structure}
\end{figure}

\begin{figure}[hbt!]
\centering
    \includegraphics[height=4in]{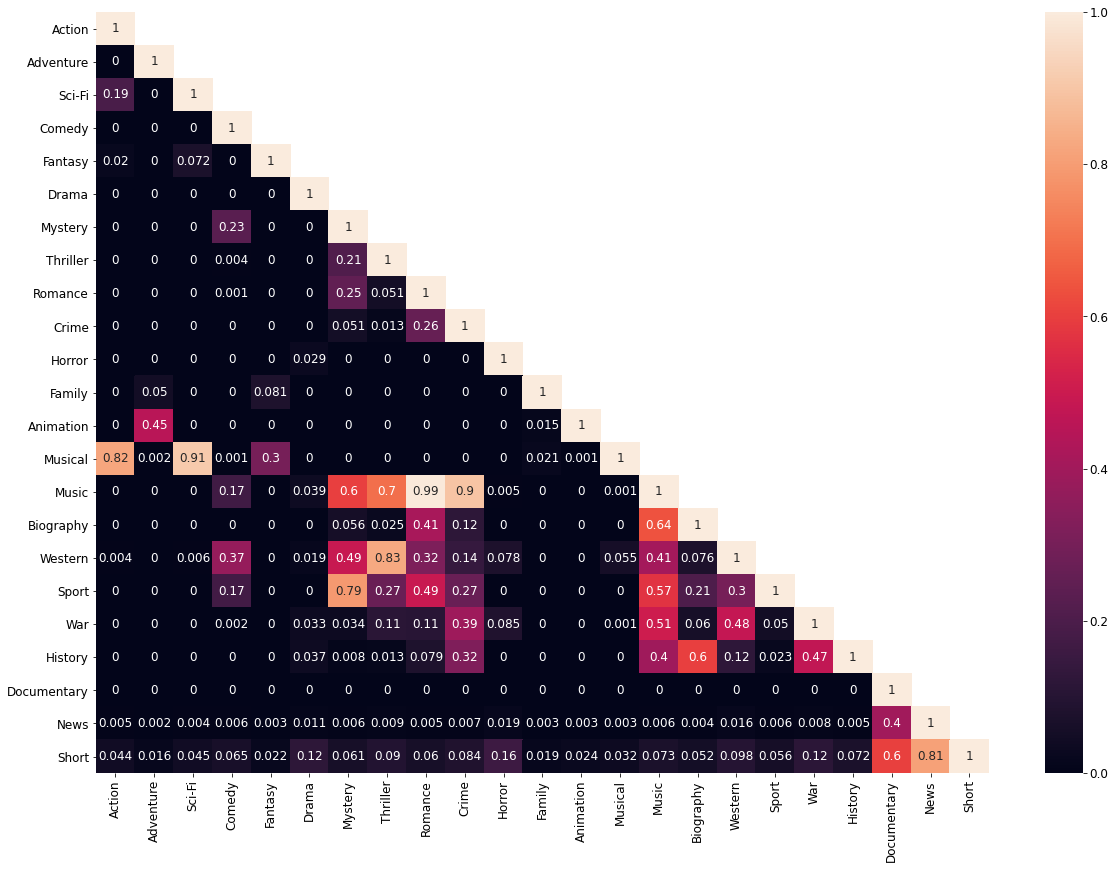}
    \vspace{1em}
    \caption{Inter genre p-value heatmap from KS test}
    \label{fig: genre_ks}
\end{figure}

\end{document}